\documentclass[referee,usenatbib]{mn2e}
\usepackage[dvipdfm]{graphicx}
\usepackage[english]{babel}
\usepackage{amsmath}

\title[Radio detection of four LBVN in the LMC]{Radio detection of nebulae around four LBV stars in the LMC}
\author[C. Agliozzo et al.]{C.~Agliozzo,$^{1,2}$\thanks{E-mail:
claudia.agliozzo@oact.inaf.it.} 
G.~Umana,$^2$ C.~Trigilio,$^{2}$ C.~Buemi,$^{2}$ P.~Leto,$^{2}$ A.~Ingallinera,$^{1,2}$
\newauthor
T.~Franzen,$^{3}$ A.~Noriega-Crespo$^{4}$\\
$^{1}$Dipartimento di Fisica e Astronomia, Sezione Astrofisica, Universit\`{a} degli studi di Catania, via S. Sofia 78, 95123, Catania, Italy\\
$^{2}$INAF Osservatorio Astrofisico di Catania, via S. Sofia 78, 95123, Catania, Italy\\
$^{3}$CSIRO Astronomy and Space Science, PO Box 76, Epping, NSW 1710, Australia\\
$^{4}$Infrared Processing and Analysis Center, California Institute of Technology, Pasadena, CA 91125, USA
}

\begin{document}

\date{}

\pagerange{\pageref{firstpage}--\pageref{lastpage}} \pubyear{2011}

\maketitle

\label{firstpage}

\begin{abstract}
The nebulae associated to four Luminous Blue Variables (LBVs) in the Large Magellanic Cloud  (LMC) 
have been observed  at 5.5 and 9 GHz using the Australia Telescope Compact Array, and radio emission has been detected 
for first time in these sources, R127, R143, S61 and S119. The radio maps of the nebulae have an angular resolution 
of $\sim 1.5"$ and a sensitivity of \hbox{1.5-3.0$\times$10$^{-2}$\,mJy beam$^{-1}$}, and show a very similar morphology 
to that observed in H$_{\alpha}$. This similarity permit us to assume that the H$_{\alpha}$ emission is not affected by strong intrinsic
 extinction due to dust within the nebulae. We estimate the masses of the ionized gas in the LBVs nebulae and their values are
 consistent with those measured in Galactic LBVs.
\end{abstract}

\begin{keywords}
stars: evolution - stars: mass-loss - ISM: bubbles
\end{keywords}

\section{Introduction}
The extended nebulae, often observed around Luminous Blue Variable stars (LBVs), constitute a quite typical characteristic of such evolved, massive  and unstable stars. 

A census of known LBVs reports 35  objects in our Galaxy and 25  in the Magellanic Clouds \citep{Clark2005, vanGenderen2001}.
Hence our knowledge of LBV nebulae (LBVNe) is still very limited due in part to the small number of known LBVs and in part to the short lifetime ($t\sim10^{-5}-10^{-4}\hspace{1mm}yr$) of the LBV phase. The characterization of LBVNe is therefore fundamental to understand the evolution of massive stars, providing estimates of important physical properties such as their mass-loss history (through the determination of the
kinematical age of the nebula),  the origin and morphology of LBVNe, and the total mass lost during the LBV phase. 

Some Galactic LBVNe have been recently studied following a multi-wavelength approach \citep{2011UmanaBulletin}, aimed at tracing both the gas and the dust components coexisting tipically in LBVNe. Among the most interesting aspects highlighted in these studies, there are: the presence of multiple shells, which are evidence of different mass-loss episodes experienced by the star \citep[e.g. G79.29+0.46,][]{2010Esteban,2011UmanaG79}; the distribution of the gas and the dust, sometimes differently shaped \citep[IRAS 18576+034,][]{2010Buemi}; the chemistry in the nebula, sometimes rich of complex molecules, such as Polycyclic Aromatic Hydrocarbons (PAHs) \citep[e.g. HD 168625,][]{2010Umana}, evidence that dust can survive despite the hostile environment due to the UV radiation from the star. 

The main mechanisms which are believed to lead to the formation of such nebulae are the moderate ($\dot{M}\sim10^{-7}-10^{-5}\hspace{1mm}M_{\odot}\hspace{1mm}yr^{-1}$) mass-loss associated to steady stellar winds   and, especially, 
the extreme mass-loss rates ($\dot{M}\sim10^{-5}-10^{-4}\hspace{1mm}M_{\odot}\hspace{1mm}yr^{-1}$) experienced by the star  during  giant eruptions, which characterize its post main sequence evolution.
\citet{1994HD} and later \citet{2006S&O} have suggested that eruptive episodes, which must form the bulk of a LBVN, are metallicity independent. If this is true, LBVs may have had a role in the evolution of the early Universe, when massive stars would have been more numerous than the present epoch, and may have provided processed material and dust for future generations of stars. 

A way to explore this possibility is to study the LBV phenomenon in environments different from the Galactic one, i.e. in galaxies with  different metallicity. Considering the capabilities of the instruments up to now available, we have started a study of LBVNe in the Large Magellanic Cloud (LMC), which is the nearest galaxy (with an accepted  distance of $D\sim48.5\hspace{1mm}kpc$) and with half its solar metallicity ($Z\sim0.5\hspace{1mm}Z_{\odot}$). 

Evidence of extended nebulae around LBVs and candidates LBV (LBVc) in the LMC has previously reported. \citet{Weis2003} and \citet{WDB2003} performed high resolution observations of 9 LBVs and LBVc in the LMC  with the H${_\alpha}$ filter
using the \emph{Wide Field Planetary Camera 2} (WFPC2)  and  ESO Multi-Mode Instrument (EMMI). In particular, they found that 5 of these objects show a very well defined shell in H$_{\alpha}$, with sizes in the range of $5"-18"$, indicating the presence of a nebula around them. 

Based on these studies we have selected a group of LVBNe that could be detected at radio wavelengths (which allow us to probe the properties of the ionized gas) through the estimation of the radio free-free emission from the observed Hydrogen H${_\alpha}$ recombination line. Here we present the first radio observations of a small subsample of these LBVNe, namely: LBVcs S61 and S119, plus LBVs R127 and R143. 

The paper is organized as follows: after an explanation of the observation and data reduction strategies ($\S$2), we compare the radio free-free continuum emission with that of H${_\alpha}$ from the morphological and photometric points of view ($\S$3 and 4). In $\S$4 we also estimate some physical parameters related to the nebulae and  in $\S$5 we present our final remarks.

\section[]{Observations and data reduction}
\subsection{ATCA observations}

Radio continuum observations were carried out on April 18 and 20, 2011, at the Australia Telescope Compact Array (ATCA), using the interferometer in the 6km-configuration. Data were acquired with the new backend system CABB, with the maximum bandwidth of 2-GHz, in the 3+6cm band.
The observations consisted of 15-min scans on the target, preceded and followed by 2-min scans on the phase calibrator (0530-727), for a total of $\sim$10 hrs  on-source integration time.  In  order to achieve the best u-v coverage, the scans were distributed over 12 values of hour angles. We used 1934-638 for the bandpass and flux calibration. Table \ref{tab:obs} summarizes the observations. 
               \begin{table}
                  \centering
                  \caption{Observational summary.}
                  \label{tab:obs}
                   \begin{tabular}{@{}lcccrrrr@{}}
                      \hline\hline
                      Observation&ATCA&Band&\multicolumn{4}{c}{Total integration}\\
                        Date& Config.&   (cm)            & \multicolumn{4}{c}{time (min)}                \\
                          &&&&&&\\
                            &               &                  & S61& R127& R143& S119\\
                       \hline
                       11 Apr 18 &6-km      &  3+6                 &            534 & 534&58&59  \\ 
                        11 Apr 21 &6-km      &  3+6                 &            58&59&533&536  \\                  
                       \hline
                    \end{tabular}
                \end{table}

Datasets were separately edited and reduced by using the Miriad software package \citep{1995miriad}. 
After applying the bandpass and time-based gain corrections, the calibrated visibilities were imported in the CASA package\footnote{\emph{Common Astronomy Software Application}, version 3.0.2.} and processed for imaging, using the task \texttt{CLEAN} and performing a natural weighting to achieve the  highest sensitivity. Dirty images were deconvolved using the Clark algorithm \citep{1980Clark}. The resulting synthetic beam $\Theta_{syn}$  is typically 2.5" $\times$ 2.0" at 5.5 GHz and 1.5" $\times$ 1.2" at 9 GHz.   For imaging R143, which is located in a very crowded region, we applied a Briggs-type weighting (parameter \texttt{robust=0}) to minimize sidelobes contribution of the brightest sources. This choice represents a good compromise between the highest resolution and sensitivity which are achievable at ATCA.

The mean noise obtained on the maps is typically \hbox{1.5-2.0$\times$10$^{-2}$\,mJy beam$^{-1}$}, consistent with the sensitivity  of the new broadband backend system CABB at these frequencies for the used integration time.

\subsection{HST data}
In order to compare the radio morphology of the nebula with that observed at other wavelengths, H$_{\alpha}$ images have 
been retrieved from the STScI data archive. These images have been obtained\footnote{Proposal ID: 6540; P.I. Regina Schulte-Ladbeck.} with the WFPC2 instrument,  using the 
H$_{\alpha}$-equivalent filter F656N, and reduced by the standard HST pipeline. These data were already published by \citet{Weis2003} and \citet{WDB2003}. For each source we combined the 
 datasets (4 images with a 500 s exposure), following a standard procedure in IRAF, to remove cosmic-ray artifacts and to 
 improve the S/N.

Finally, we have also astrometrically recalibrated the HST images using the NOMAD catalog \citep{2005Nomad}, for a corrected overlay with the
 radio images.

\section{Morphology of the radio nebulae}

\begin{figure*}
 \vbox to220mm{\vfill 
    \includegraphics[scale=1.3,angle=90]{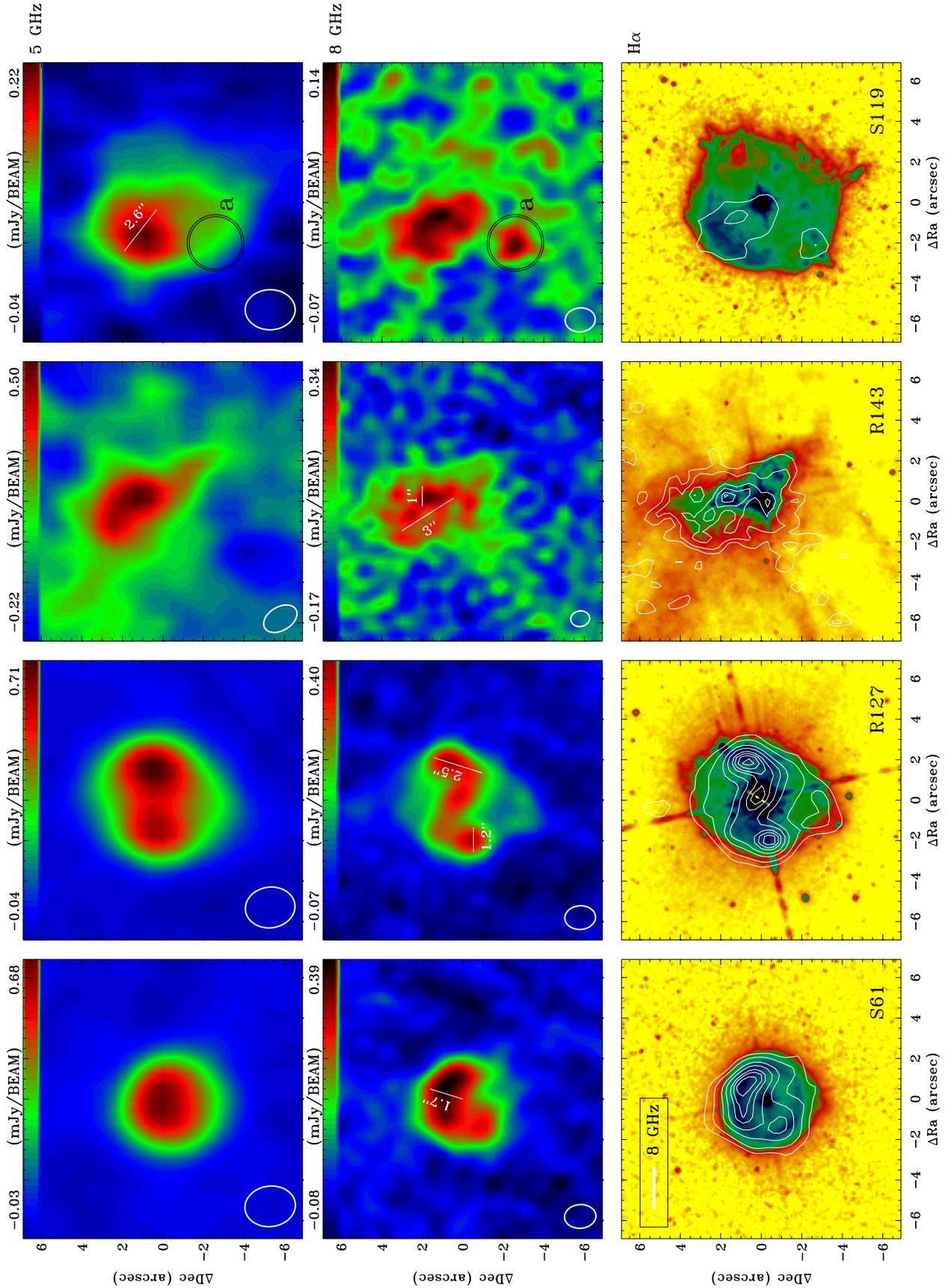}
   \caption{Top panels: 5 GHz ATCA maps of S61, R127, R143 and S119 from the left to the right. Middle panels: 8 GHz ATCA maps in the same order as before. Bottom panels: H$_{\alpha}$ WFPC2/HST images overlayed with 8 GHz ATCA contours. Contours are chosen at (3, 5, 8, 10, 12, 13, 14, 15)$\sigma$. In the radio maps the white ellipses represent the synthetic beam.}
   \vfill
\label{landfig}

   }
\end{figure*}

                \begin{table}
                  
                  \caption{Observed flux densities S$_{obs}(\nu)$ at 5.5 and 9 GHz, angular sizes at 9 GHz.}
                   \begin{center}
                  \label{tab:fluxobs}

                   \begin{tabular}{lccc}

                      \hline\hline
                      Source&S$_{obs}$(5.5 GHz)&S$_{obs}$(9 GHz)& Angular \\
                            &    (mJy)     & (mJy)   &sizes\\
                       \hline
                       S61  &     2.1  $\pm$ 0.1        &  2.2  $\pm$ 0.3     &5.1"$\times$5.7"\\ 
                       R127 &     3.1 $\pm$ 0.2         &    3.3$\pm$ 0.4   &6.6"$\times$7.5"\\  
                       R143&  2.4$\pm$ 0.9&4.0$\pm$ 0.7& 3.7"$\times$3.2"\\                           
                       S119 &1.0 $\pm$ 0.2 &  0.9 $\pm$ 0.2  &8.0"$\times$9.6"$^{a}$ \\   
                       \hline
                    \end{tabular}
                    \end{center}
                     $^{a}$ From the map at 5.5 GHz.\\
                \end{table}
                  
In this section we give a morphological description of the radio emission compared to the optical one \citep[already provided
 by][]{Weis2003, WDB2003}, obviously limited by the lower resolution of the radio maps. Fig. \ref{landfig} shows all the maps which we obtained in this work, compared with the reprocessed HST images.

The angular sizes of the emission region were estimated drawing a box above $3\sigma$ on the sources. The maximum and minimum sizes are hence reported in Table \ref{tab:fluxobs}. The estimated values are consistent with the precedent measurements \citep[e.g.][]
{1993Clampin, 1998Smith, 1999Pasquali, WDB2003, Weis2003}. 

\paragraph*{S61} is a LBV candidate \citep{1977Walborn,1982Walborn}, with a H$_{\alpha}$ nebula consisting of an inner 
brighter ring-like structure and a larger diffuse emission \citep{Weis2003}. The radio nebula here resembles the H$_{\alpha}$ 
one, being roughly spherical (Fig. \ref{landfig}). The nebula is fainter in the south-west and the emission is not homogeneous: 
the brightest part is in the north-east, where the gas might be  thicker. A larger diffuse emission all around is evident above 3$
\sigma$, as well as the outer shell seen in the H$_{\alpha}$.

\paragraph*{R127}
is classified as LBV star, after it showed a S-Dor type variability \citep{1983Stahl}. \citet{1993Clampin} reported the presence of a diamond shaped optical nebula and then the high resolution HST image in \citet{Weis2003} revealed a nearly spherical structure in the central region and a fainter emission in the north and southern direction. The shell presents also two brighter rims on the east and the north-west side.

The radio nebula show two components: a thin roughly spherical nebula, elongated in the N-S direction; a thicker bipolar nebula, Z-pattern shaped and centered in the optical stellar position. A hint at the central object is visible in the radio image. 


\paragraph*{R143}
is a confirmed LBV \citep{1993Parker}, located in the 30 Doradus cluster. Around the star, there is a nebula, elongated in the North-South direction, triangular shaped, detected with the HST \citep{1998Smith,Weis2003}. As pointed out by \citet{Weis2003}, the nebula is concentrated to the West of the star, without a counterpart nebular emission to the East. Four curved filaments close to R143 were noticed by \citet{1961Feast} and then attributed to the nearby HII region by \citet{1998Smith}.

 The morphology in the radio map is quite similar to the optical one. The overlay with the optical image (Fig. \ref{landfig}, bottom panel) show that the nebula might be contaminated by the near 30 Dor HII region emission, as the optical one is.

An unresolved compact emission in the position of the star is also visible in the map at 9 GHz.

\paragraph*{S119}
\citet{1994Nota} proposed S119 as a LBV candidate, after the discover of a nebula around the previously classified Ofpe/WN9 star \citep{1989Bohannan}. The nebula appeared as a shell with a brighter lobe at the North-East direction \citep{1994Nota}. With the HST high resolution image, \citet{WDB2003} detected small scale structures, such as filaments and knots which give to the nebula a patchy appearance. 

The marked asymmetry of the nebula was explained \citep{2001DC} as due to a bow shock interaction with the ISM, since the S119 radial velocity \citep[blue-shifted $\sim$100 $km\hspace{1mm}s^{-1}$ from the LMC systemic velocity,][]{1994Nota} suggests a runaway nature for the star.

The diffuse emission visible at 6 cm above 3$\sigma$ (green/blue levels in Fig. \ref{landfig}, top panel) is not detected at ATCA 3 cm, but the thicker N-E and S-E parts are well detected and resemble the H$_{\alpha}$ morphology. In particular the S-E compact region (circle "a" in the figure) is not visible at the lower frequency. This might be due to the presence of an optical thick HII region. We can speculate that this is the result of an interaction between the stellar ejecta and the close ISM.               

\section[]{Data Analysis}

              \begin{table}
                   \begin{center}
                   \caption{Electron density n$_{e}$, electron temperature T$_{e}$, effective temperature T$_{eff}$ available in the literature.}

                  \label{tab:literature}
                   \begin{tabular}{lccc}

                      \hline\hline
                      Source&n$_{e}$&T$_{e}$&T$_{eff}$\\
                         &(cm$^{-3}$)&(K) & (K)\\
                          &&&\\
                           
                       \hline
                       S61  & 400$^{a}$&6120$^{a}$& 27600$^{c}$\\ 
                       R127 & 720$^{b}$&6420$\pm$ 300$^{b}$ &16000$^{d}$\\  
                       R143& 1000$^{b}$&12200$\pm$ 1500$^{b}$ &9700$^{e}$\\                           
                       S119 & 680$^{b}$&$<$6800$^{b}$ &26200$^{c}$\\   
                       \hline
                    \end{tabular}
                     \end{center}
                     $a$ Pasquali et al. 1999\\
                     $b$ Smith et al. 1998\\
                     $c$ Crowther \& Smith 1997\\ 
                     $d$ Stahl et al. 1983\\
                     $e$ Davies et al. 2005\\
             \end{table}

                 \begin{table}

                  \caption{ H$_{\alpha}$ line integrated flux F(H$_{\alpha}$), expected free-free density flux S$_{exp}(\nu)$.}
                  \label{tab:fluxexp}

                  \begin{center}
                   \begin{tabular}{lccc}
                 
                      \hline\hline
                      Source&F(H$_{\alpha}$)&S$_{exp}$(5.5 GHz)& S$_{exp}$(9 GHz)\\
                            &(\hbox{10$^{-14}$\,erg cm$^{-2}$ s$^{-1}$})&(mJy)   &   (mJy)\\
                       \hline
                       S61  &10.6$\pm$0.7& 1.7$\pm$0.1  &1.6 $\pm$0.1\\ 
                       R127 & 20.0$\pm$1.7  &3.3$\pm$0.3  & 3.2$\pm$0.3\\  
                       R143& 3.6$\pm$2.1  &0.84$\pm$0.50&0.80$\pm$0.48\\                           
                       S119 & 6.7$\pm$0.7   &1.14 $\pm$0.11   &1.09$\pm$0.11\\   
                       \hline
                    \end{tabular}
                     \end{center}

                \label{derived}
                \end{table}

\subsection{Intrinsic extinction and spectral index}
Under typical conditions of the ISM, the optical emission can be affected by extinction due to dust, while radio emission 
does not. We can establish if the optical emission of ionized nebulae is affected by intrinsic extinction, following one of the methods for measuring extinction described by \citet{1984Pottasch}. It consists in the comparison between the recombination hydrogen line and the radio continuum emission coefficients (having the same dependence on the nebular density), through the following relation
\begin{equation}
 S_{\nu} = 2.51\times10^{7}T_{e}^{0.53}\nu^{-0.1}YF(H_{\beta})_{exp} \hspace{0.5cm}[Jy]
\label{predicted}
\end{equation}
where $T_{e}$ is the electron temperature of the nebula in K unit,
 $\nu$             is the radio frequency in GHz,
 Y is a factor incorporating the ionized He/H ratios
 \begin{equation}Y=1+\frac{n(He^{+})}{n(H^{+})}+4\frac{n(He^{++})}{n(H^{+})}\frac{ln(4.95\times10^{-2}T_{e}^{\frac{3}{2}}\nu^{-1})}{ln(9.9\times10^{-2}T_{e}^{\frac{3}{2}}\nu^{-1})}\label{Y}\end{equation} and $F(H_{\beta})_{exp}$ is the expected H$_{\beta}$ flux when the nebula is not affected by intrinsic extinction. For a nebula with electron density less than $10^{4} cm^{-3}$ and electron temperature $\sim 10^{4} K$, $F(H_{\beta})_{exp}=\frac{F(H_{\alpha})_{exp}}{2.85}$ \citep[see Table III-1 in][]{ 1984Pottasch}. 
To evaluate the second term in equation (\ref{Y}), we consider that in a low ionization nebula (see $T_{eff}$ in Table \ref{tab:literature}), where the ionization is due mainly to the radiation field of the star, and with small $\frac{n(He)}{n(H)}$ ratios \citep[typically 0.1-0.2 in the nebula, e.g.][]{2001Lamers}, the contribution from  $n(He^{+})$ to the free-free emission is negligible, consequently also that one from $n(He^{++})$.

Hence from the HST images we have measured the flux emitted in the H$_{\alpha}$. We have extracted aperture photometry fitting a circle around the star and another around the nebula. For R143, which does not have a rounded shape, we used a polygonal line to select the region where  to measure the flux density. Since we are interested in the recombination line emission of the ionized nebulae, we have substituted the stellar contribution with the mean brightness around the star. In the case of R127 we have also subtracted the contribution from the external star which is projected on the nebula. Fluxes in the H$_{\alpha}$ are reported in Table \ref{tab:fluxexp}.

We have measured the radio flux densities integrating over the entire nebula by using an interactive tool of CASA. The rms noise $\sigma$ has been evaluated far from the targets. Flux densities were therefore compared with the sum of the clean components inside the regions. Being both the measures consistent each other, we report the values measured with the interactive mode in Table \ref{tab:fluxobs}. Flux density errors are defined as $\epsilon=\sigma \sqrt{N}$, where $N$ is the number of independent beams in the selected region \citep{2006Brogan}. 
The error due to calibration uncertainty\footnote{$<$10\% for 1934-638.} is not taken into account.

Finally, with Y=1 and using as $T_ {e}$ the values found in the literature (Table \ref{tab:literature}), from the measured H$_{\alpha}$ we have estimated the radio fluxes at the observed frequencies (Table 4) and compared to the observed ones (Table \ref{tab:fluxobs}). 

The angular resolution obtained is not sufficient to isolate the stellar objects. Evidence of stellar wind is the positive value of the spectral indeces $\alpha$ (Table \ref{tab:mass}), which indicate a contribution to the nebular free-free optically thin emission, more important at higher frequencies. Despite this, the values that we found indicate that the H$_{\alpha}$ emission does not probably suffer from strong intrinsic extinction. This is apparently true for both R127 and S119, whose density fluxes are equal to the expected ones within the errors. Moreover for R127 we have evidence of a stellar wind contribution at 9 GHz. In the case of S119 we cannot assess it because of the sensitivity limit at 9 GHz.

A similar discussion on the central object contribution can be applied to S61 and R143, even if in the last case the spectral index evidently shows that the nebula flux is contaminated by other emission sources (i.e. 30 Dor HII region). Moreover we cannot exclude the presence of dust in the nearby, which may cause optical extinction.

The S61 flux density at both 5.5 and 9 GHz are higher than that ones expected from the recombination line flux estimation. This could be due to an important stellar wind contribution even at the lowest frequencies or to a non-negligible dusty nebular component.

               \begin{table*}

                   \begin{center}
                    \caption{Spectral index $\alpha_{obs}$, Emission Measure $<EM>$, Linear size, geometrical depth $s$, source solid angle $\Omega_{s}$, average electron density $<n_{e}>$, ionized mass $M_{ionized}$.}
                    \label{tab:mass}
                   \begin{tabular}{lccccccc}

                      \hline\hline
                      Source&$\alpha_{obs}$&$<EM>$&Linear&$s$&$\Omega_{s}$ &$<n_{e}>$&$M_{ionized}$\\
                            && $[pc\hspace{1mm}cm^{-6}]$&size (pc$\times$pc) &$(pc)$&(arcsec)$^{2}$ &$(cm^{-3})$&$(M_{\odot})$\\
                       \hline
                       S61  &0.1& 1364  &1.2$\times$1.3& 0.40& 24.5& 58 & 0.78  \\ 
                       R127 &0.1&1345  &1.6$\times$1.8& 0.28-0.59& 38.1& 69-48 & 1.01-1.46 \\  
                       R143& 1.0&1697 & 0.9$\times$0.8& 0.24-0.71& 19& 85-49 & 0.52-0.90\\                           
                       S119 &-0.2 $^{a}$& 222& 1.9$\times$2.3 & 0.61& 71& 19 & 1.13 \\   
                       \hline
                    \end{tabular}
                    \end{center}
                    $^{a}$ This value must be considered a lower limit.\\
                \end{table*}
\subsection{Mass of the ionized nebula}
From the radio free-free emission measurement we can estimate the total ionized mass as the total particles present in the nebula. For a not self-absorbed free-free emission, the electron density in a nebula can be determined through the relation between the Emission Measure 
\begin{equation}
EM=\int_{0}^{s}{n_{e}^{2}dl\hspace{1mm}[pc\hspace{1mm}cm^{-6}]}
\label{equ:em}
\end{equation}
 and the optical depth $\tau_{ff}$
\begin{equation}
\tau_{ff}=8.24\times10^{-2}T_{e}^{-1.35}\left(\frac{\nu_{[GHz]}}{5}\right)^{-2.1}EM
\end{equation}
where $\tau_{ff}$ is evaluated for each pixel from the observed brightness $B_{\nu}$ which, in the optical thin case is   $B_{\nu}=B(T)\tau$, where $B(T)$ is a black body with an assumed temperature T equal to the gas temperature $T_{e}$ (see Table \ref{tab:literature}).

From the mean $<EM>$, measured in the map, inverting equation \ref{equ:em}, it is possible to estimate a mean value for the particle density $<n_{e}>$. Assuming that the nebulae are uniform, the emission measure is simply $EM=<n_{e}>^{2}s$, where the geometrical depth $s$ needs a strong hypothesis on the nebular geometries. 

It has been believed that LBVNe are often torus-like shaped, indicating that mass-loss happens in a preferred direction \citep[e.g.][]{2003Clark, 2005Davies}. If it is true, the nebulae associated to R127, S61 and S119 may be torus seen pole-on, and that one of R143 seen roughly edge-on. Hence we assume a depth equal to the transversal size measured on the maps, chosen where the emission is the $70\%$ of the peak flux density (as shown with white lines in Fig. \ref{landfig}). In the cases of R127 and R143 we calculated the electron density and the ionized mass assuming two different geometrical depths (1.2" and 2.5" for R127, 1" and 3" for R143).

Finally the total ionized mass is given by the total particles in the nebula, i.e. by $M_{ionized}=<n_{e}>\frac{m_{p}}{M_{\odot}}V$. The volume assumed for each nebula is given by the area where we have estimated the $<EM>$ and the assumed geometrical depths. The resulting masses are reported in Table \ref{tab:mass}, together with the other physical parameters. The linear sizes in the table are obtained assuming for the nebulae the distance of the LMC ($D=48.5\hspace{1mm} kpc$).

From our estimation, the masses range between $0.5$ and $1.5\hspace{1mm}M_{\odot}$, which are typical values for nebular masses of the Galactic LBVs \citep[in general few solar masses,][]{1994HD, Clark2005}.

\section{Discussion/Conclusions}

The improved sensitivity at ATCA with the new CABB correlator has allowed us to obtain the first radio detections of the nebula around four LBV type stars in the Large Magellanic Cloud. Even if our highest resolution cannot distinguish angular scales smaller than 1.5" (that means structures of linear sizes $\sim$0.35 pc at a distance of 48.5 kpc), it was possible to measure the nebular sizes and to compare the radio morphology with the optical one. The linear sizes are consistent with previous estimates at different wavelengths \citep[H$_{\alpha}$,][]{Weis2003, WDB2003} as well as with the Galactic examples \citep[e.g.][]{2003Clark}. No evident differences are visible, as expected if the $H_{\alpha}$ is not intrinsically extincted. On the contrary, the radio resembles the optical.
A similar result was found by \citet{2002DuncanWhite}, studying four Galactic LBVNe and comparing the radio images with the H$_{\alpha}$ ones. The strict similarity was particularly evident in AG Car.

The radio flux densities were compared here with the extrapolated fluxes from the recombination
hydrogen line. The HST H$_{\alpha}$-equivalent filter F656N has a bandwidth $\approx 28\hspace{1mm}
\AA$, with the  
H$_{\alpha}$ line in the band center, and does not cover H$_{\alpha}$ velocity features higher in magnitude
than $640\hspace{1mm}km\hspace{1mm}s^{-1}$. However high-resolution Echelle spectra
in \citet{Weis2003} showed that high velocity outflows are not present in such nebulae.
On the other hand, the filter can detect signal from [NII] 6548 \AA \hspace{1mm}  
emission. Assuming a mean value for the [NII]6584 \AA/H$_{\alpha}$ \citep[0.6,][]{1998Smith} and that this is
valid also for the [NII] 6548 \AA line, and considering that the filter response at 6548 \AA \hspace{1mm} is about the 25\% of the
maximum, then the H$_{\alpha}$ emission evaluated here may be overestimated of the 15\%.
And this has to be considered an upper limit.

Such contamination of H$_{\alpha}$ due to [NII] is, however, within the flux errors,
for most of our sources. If we assume that all the flux falling in the filter is due to  H$_{\alpha}$,
the comparison between the radio and optical fluxes does not show any differences and this lead us to
suppose
that there is not intrinsic extinction,
especially in the case of R127 and S119. On the other hand,  if there were some extinction,
this  will be attributable to uniformly distributed dust, as the radio morphology
resembles the optical one and we may exclude any clumping in the dust distribution.
 
A similar discussion must be done with caution in the case of S61, whose optical emission could be affected by intrinsic extinction due to dust, and in the case of R143, which is probably affected by the emission of the close HII region, as well as of a likely dusty environment.

The  derived spectral indeces indicate a possible contribution from a central object that must be considered and, more specifically, that current mass loss from the central star is ongoing.

In addition, values for the ionized mass have been obtained in this work. Estimates for the nebular mass of S119 and R127 have been previously reported. In particular Nota et al. 1994  and \citet{1993Clampin} provided a mass of $M=1.7\hspace{1mm}M_{\odot}$ for the nebula of S119 and of $M=3.1\hspace{1mm}M_{\odot}$ for R127. These results were obtained integrating H$_{\alpha}$ emission luminosity and adopting values for the electron density of the order $800-1000\hspace{1mm}cm^{-3}$, derived by [SII 6716/6731] \AA \hspace{1mm} line ratio, and a temperature of 7500 K. A further value for the mass of R127 was found by \citet{2009Munari} through a photoionization modeling of the emission line spectrum in the range $8400-8800$ \AA. They found a mass of $1.33\times10^{-3}\hspace{1mm}M_{\odot}$. 

We are conscious that the ionized mass estimated with this method is very sensible to the assumed geometry and that we are not taking into account the filling factor correction. If these circumstellar nebulae are not homogeneous as we assumed, the true average density, and hence the nebular masses, are smaller than the estimates provided here by the value of the filling factor \citep[tipically between $\sim0.2-0.7$ for extended nebulae, e.g.][]{1988Mallik,1994Boffi}.
Both these facts may lead to the discrepancy with the measurements reported in \citet[Table 9,][]{2003Clark}.

The contribution from the stellar wind to the total emission can be neglected. Infact assuming a mass-loss rate of $\dot{M}\sim10^{-5}\hspace{1mm}M_{\odot}\hspace{1mm}yr^{-1}$ and a stellar wind velocity of $v_{\infty}\sim100\hspace{1mm}km\hspace{1mm}s^{-1}$, which are typical values for Galactic LBVs, the flux density due to the stellar wind will be of the order of 0.03 mJy at 9 GHz \citep{1975PF}, i.e. only the $1\%$ of the observed flux density. This means that the overestimation of the mean electron density and of the ionized mass are negligible. In the case of R143 an overestimation could be assigned to the close HII region emission.
The values reported here must, then, considered only as indicative, but are almost consistent with the values in \citet{1993Clampin} and \citet{1994Nota}.

Thanks to its properties, radio emission can be used as a good probe of the ionized gas emission from a nebula around a hot star. This is important when one wants to determine the gas content to establish the total mass budget ejected during the LBV phase and to constrain stellar evolution models of very massive stars. For the moment, the resolution achieved at ATCA in this work doesn't allow us to separate the nebular and stellar contributions to the total emission.

In addition to these, new higher resolution observations can provide a more detailed morphological study of the nebula, as well as a possibly detection of the central stellar wind and hence the current mass loss rate, where instead the optical H$_{\alpha}$ estimates can be uncertain.

\section*{Acknowledgments}

The Australia Telescope Compact Array is part of the Australia Telescope National Facility which is funded by the Commonwealth of Australia for operation as a National Facility managed by CSIRO. This research has made use of the SIMBAD database, operated at CDS, Strasbourg, France. It is also based on observations made with the NASA/ESA Hubble Space Telescope, and obtained from the Hubble Legacy Archive, which is a collaboration between the Space Telescope Science Institute (STScI/NASA), the Space Telescope European Coordinating Facility (ST-ECF/ESA) and the Canadian Astronomy Data Centre (CADC/NRC/CSA).

\bsp

\label{lastpage}

\end{document}